\documentclass[prd, aps, superscriptaddress, preprintnumbers,
floatfix, nofootinbib]{revtex4}

\usepackage{amsfonts} \usepackage{amsmath} \usepackage{amssymb}
\usepackage{bm} \usepackage{dcolumn} \usepackage{epsfig}
\usepackage{graphicx} \usepackage{graphics}
\usepackage[latin1]{inputenc} \usepackage{latexsym}
\usepackage{rotating} \usepackage{hyperref}
\usepackage{color}
\usepackage{amsfonts} \usepackage{amsmath} \usepackage{amssymb}
\usepackage{bm} \usepackage{dcolumn} \usepackage{epsfig}
\usepackage{graphicx} \usepackage{graphics}
\usepackage[latin1]{inputenc} \usepackage{latexsym}
\usepackage{rotating} \usepackage{hyperref}

\newcommand\be{\begin{equation}}
\newcommand\bea{\begin{eqnarray}}
\newcommand\ee{\end{equation}}
\newcommand\eea{\end{eqnarray}}

\begin{document}

  \title[Hints from Superstring Theory]
    {Beyond Standard Inflationary Cosmology}

  \author{Robert H. Brandenberger\\[1\baselineskip]
    Physics Department, McGill University\\
    3600 University Street, Montreal, QC, H3A 2T8\\
    Canada}

  

\begin{abstract}

The inflationary scenario is not the only paradigm of  early universe cosmology which is
consistent with current observations. General criteria will be presented which any successful
early universe model must satisfy. Various ways, including inflation, will be presented which
satisfy these conditions. It will then be argued that if nature is described at a fundamental level
by superstring theory, a cosmology without an initial space-time singularity
will emerge, and a structure formation scenario
which does not include inflation may be realized. 

\end{abstract}


  \maketitle

  
  \section{Introduction}
  
  \subsection{Goals of Early Universe Cosmology}
  
  Explaining the origin and early evolution of the universe
  has been a goal of cosmology for millenia. However, it is only
  over the past few decades that cosmology has moved from being
  a branch of philosophy and theology to being a mainstream
  area of physics research. This change is due to an explosion of data about
  the structure of our universe which experimentalists and observers
  have gathered. 
  
\begin{figure}
    \includegraphics[scale=0.4]{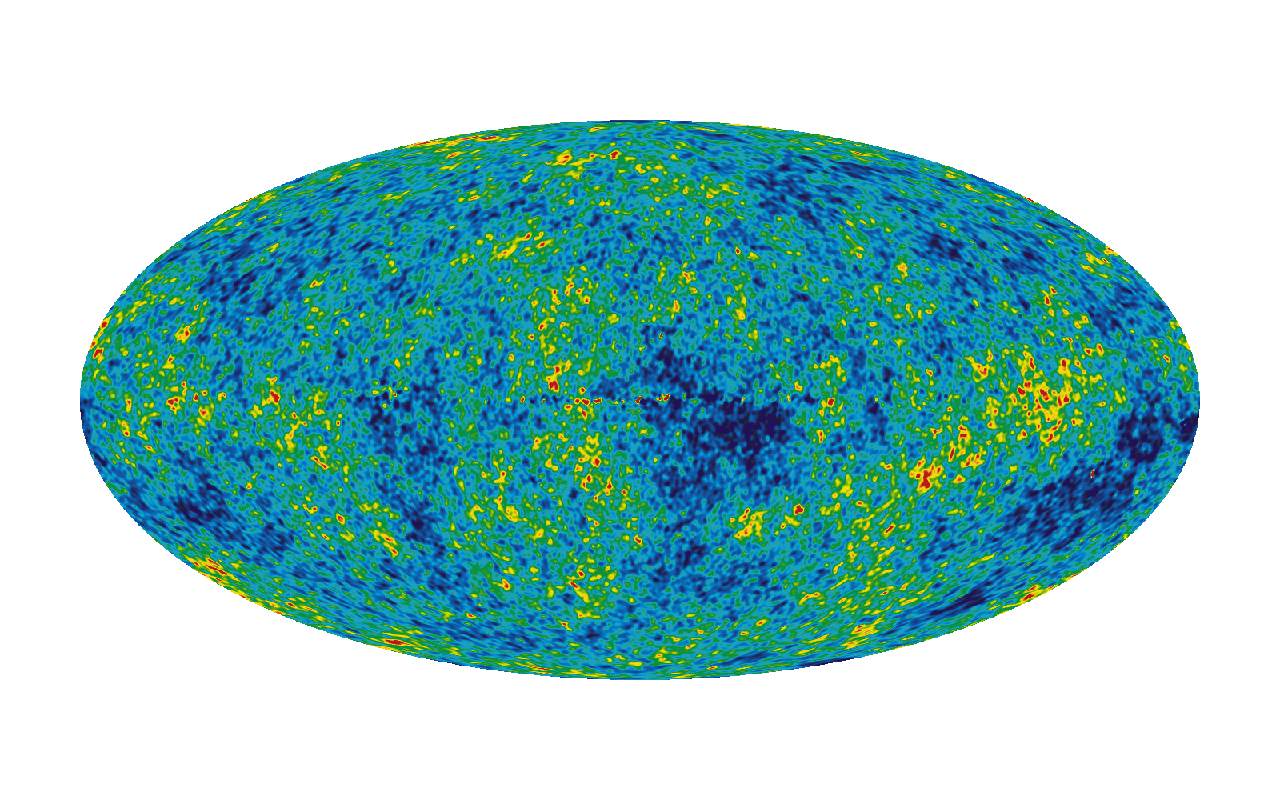}
    \caption{Map of the microwave background temperature anisotropies. What is shown
    is a projection of the sky onto a plane. The colour scale indicates temperature.
    The temperature differences between hot and cold areas is of the order $10^{-5}$
    of the average temperature. Credit: NASA/WMAP Science Team.}
    \label{WMAP}
  \end{figure}

\begin{figure}
    \includegraphics[scale=0.55]{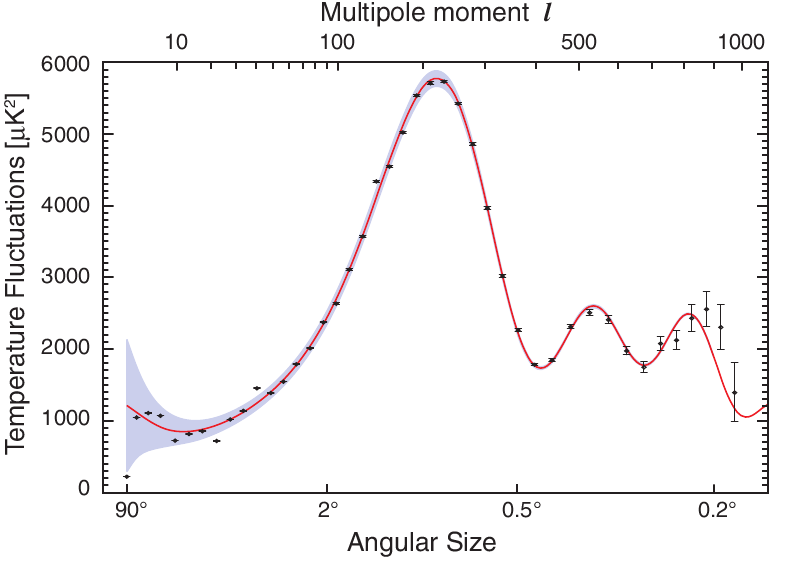}
    \caption{Angular power spectrum of the data shown in the previous
    figure. The horizontal axis is inverse angular scale. The vertical axis gives
    the mean magnitude of fluctuations on the corresponding angular scale.
    Credit: NASA/WMAP Science Team.}
    \label{WMAPspect}
  \end{figure}

  The prime example of data is the Cosmic Microwave Background (CMB).
  The CMB was discovered serendipitously in the 1960s \cite{Penzias}. Only
  in the early 1990s detailed measurements of the black body
  nature of the spectrum became available \cite{Gush, COBEA}, and 
  the first anisotropies were discovered \cite{COBEB}. Over the
  following two decades, rapidly improving measurements of the angular
  power spectrum \footnote{The CMB temperature map of the sky
  can be expanded in terms of spherical harmonics (the analog of
  Fourier expansion in Euclidean spaces). The spherical harmonics
  are labelled by two integers $l$ and $m$ with $-l \leq m \leq l$. The coeffcient of the
  $(l, m)$ spherical harmonic gives the amplitude of the CMB
  anisotropies on an angular scale proportional to $l^{-1}$ and in a
  direction given by $m$. If we average the square of the coefficients
  for fixed $l$ over the allowed values of $m$, we obtain what is called
  the {\it angular power spectrum} of CMB anisotropies.}
  of the CMB anisotropies were made, culminating with
  the results from the WMAP \cite{WMAP} \footnote{WMAP stands
  for {\it Wilkinson Microwave Anisotropy Probe}.} and Planck \cite{Planck}
  satellites.  To better than 1 part in $10^4$, the temperature
  of the CMB is isotropic. At a slightly lower level, anisotropies appear. 
  Figure \ref{WMAP} depicts the map of CMB anisotropies from the
  WMAP satellite. In this figure, the sky is projected onto a plane
  in the same way that the surface of the Earth is sometimes projected
  onto a plane. To quantify the anisotropies, we can expand the
  map in spherical harmonics and compute the angular power
  spectrum. The
  results are shown in Figure \ref{WMAPspect} where the horizontal axis is the
  value of $l$ (or equivalently the angular scale), and the vertical
  axis gives the amplitude. The figure shows several interesting
  features. First of all, there are oscillations in the spectrum with
  a first peak at an angular scale of about $1^{o}$. Secondly, at
  large angular scales the spectrum is quite flat. Finally, on small
  angular scales the spectrum is suppressed. One of the main goals
  of early universe cosmology is to provide an explanation for this
  data.
  
  The CMB is not the only window we have to probe the structure of
  the universe. Using optical and infrared telescopes we can study
  the distribution of galaxies and galaxy clusters. X-ray telescopes
  allow us to map out the distribution of hot gas (gas with a
  temperature much higher than the surface temperature of the
  sun) in the Universe. Radio
  telescopes allow the exploration of cold gas (gas with a
  temperature lower than the surface temperature of the sun).  A new window to
  probe the universe is emerging: the 21cm window \cite{Furlanetto}
  (radiation at a frequency of 21cm),
  a window which allows us to map out the distribution of neutral
  hydrogen, the atom which dominated matter content, up to very
  large distances and early times. These windows have the added
  advantage (compared to the CMB) of providing us with 
  three-dimensional maps as opposed to only two-dimensional ones
  (given a cosmological model).
  
  The second main goal of early universe cosmology is to provide
  explanations for this data based on physical models which obey
  the principles of causality. As we will see, Standard Big Bang Cosmology,
  the paradigm of cosmology until 1980, cannot provide a causal
  explanation (an explanation from a theory which obeys the
  principles of Special Relativity, namely that no information can travel
  faster than the speed of light) of the data, and we have to go back to the very early
  universe if we want to find explanations.
  
  An important aspect of cosmology as a branch of physics is that,
  once we have a model which can address the data, we can make
  predictions for future observations. As will be discussed below,
  the current paradigm of early Universe cosmology made
  predictions concerning data which was not yet available at the
  time when the model was formulated. As mentioned below,
  Standard Big Bang cosmology predicted the existence and
  black body nature of the CMB, and inflationary cosmology
  predicted an almost spatially flat universe with a spectrum of
  CMB anisotropies with specific properties. These predictions
  were confirmed much later once observations became available,
  thus ruling out various alternatives cosmological models. I believe
  that any alternative to the current cosmological paradigm
  must make specific predictions for future observations
  with which it can be distinguished from the current paradigm.
  I regard this falsifiability aspect as an important challenge
  for modern cosmology.
  
  Summarizing, the goals of early universe cosmology are:
  
  \begin{itemize}
  
  \item Explain the origin and early evolution of the universe.
  
  \item Explain the currently available data on the large-scale
  structure of the universe based on causal physics.
  
  \item Make predictions for future observations.
  
  \end{itemize}
  
  A few words on the notation. We will mostly consider homogeneous
  and isotropic space-times in which the line element is given by
  \be \label{bgcosmo}
  ds^2 \, = \, dt^2 - a(t)^2 d{\bf x}^2 \, ,
  \ee
  where $t$ is physical time (time measured by our physical
  clocks), ${\bf x}$ are comoving spatial
  coordinates (a coordinate system in which the coordinates of
  points locally at rest do not change as space
  expands), and $a(t)$ is the cosmological {\it scale factor}.
  Particles at rest have time-independent comoving coordinates,
  and the change in $a(t)$ describes the expansion (or contraction)
  of space. Light travels on curves ${\bf x}(t)$ for which $ds^2 = 0$.
  It is often useful to work with {\it conformal time} $\tau$ in terms
  of which light travels at $45^{o}$ lines when drawing space-time
  diagrams in terms of conformal time and comoving spatial coordinates.
  The Hubble expansion rate $H(t)$ is given by
  \be
  H(t) \, = \, \frac{{\dot a}}{a} \, ,
  \ee
  where an overdot represents the derivative with respect
  to time.
  
  We are using units in which the speed of light, Planck's
  constant and Boltzmann's constant are all set to $1$.
  These are units which are generally used in the field of
  high energy physics.
  
  \subsection{Problems of Standard Big Bang Cosmology}
  
  Cosmology deals with space, time and matter.
  Standard Big Bang Cosmology (SBB) is based on describing
  space and time using Einstein's classical theory of General
  Relativity, and matter in terms of classical perfect fluids. At
  the present time (which will be denoted by $t_0$ in
  the following), matter is dominated by a fluid with vanishing pressure,
  which in cosmology is called {\it cold matter}, while the CMB provides a
  component which is sub-dominant at the present time.
  
  The equations of motion of General Relativity imply that
  in the presence of homogeneously distributed matter,
  the scale factor $a(t)$ cannot be constant. Space is
  either expanding or contracting. Observations tell us
  that we live in an expanding universe. Since the energy
  density in radiation increases faster (namely as
  $a(t)^{-4}$) as we go back in time
  than the energy density in cold matter (which increases
  as $a(t)^{-3}$), there is a time $t_{eq}$ when the energy
  densities of the two components are equal. Before $t_{eq}$
  the universe is dominated by radiation, afterwards by
  cold matter. It turns out that $t_{eq}$ is the time before
  which inhomogeneities do not increase in time (up to
  factors which are logarithmic in time) \footnote{The dominant
  radiation causes a homogeneous gravitational potential
  which impedes the growth of density fluctuations in
  the cold matter.}. 
  
 There is another time which is important in the evolution of
 the late universe. This is the time $t_{rec}$ of recombination
 before which the energy density was larger than the
 ionization energy of hydrogen. Hence, for $t < t_{rec}$ the universe
 was an electromagnetic plasma, and only after which it
 becomes neutral. The photons (particles of light) which are 
 present at the time $t_{rec}$ can then travel unimpeded to us.
 This is in fact the origin of the CMB.
 
 For $t < t_{eq}$ the scale factor increases as
 \be
 a(t) \, \sim \, \bigl( \frac{t}{t_0} \bigr)^{1/2} \, ,
 \ee
 and vanishes at the time $t = 0$. At this time, the curvature
 and energy densities become infinite. This is the {\it Big Bang}
 singularity of standard cosmology. 
 
\begin{figure}
    \includegraphics[scale=0.7]{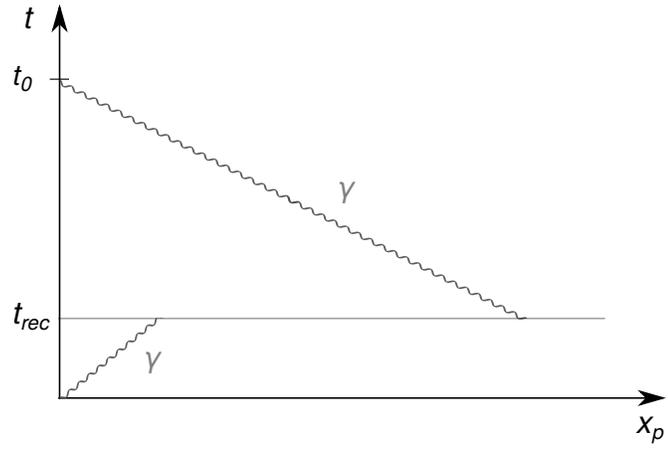}
    \caption{Illustration of the horizon problem. The vertical axis is time, the
    horizontal axis is distance - implicitly using coordinates in which light
    travels at $45$ degree angles. We see light rays (wavy lines) starting at
    the time $t_{rec}$ of recombination and reaching us today (time $t_0$).
    As shown, the distance over which we see the microwave light to be
    isotropic is much larger than the distance which could have been in
    causal contact starting at the initial time $t = 0$.}
    \label{horizon}
  \end{figure}

 The most impressive success of Standard Big Bang Cosmology
 is that it predicted the existence and black body nature of the CMB.
 However, the scenario is not able to explain the near isotropy
 of this radiation. This is the so-called {\it horizon problem} and it
 is illustrated in Figure \ref{horizon}. Here, the horizontal axis indicates comoving
 spatial coordinates, the vertical axis in conformal time.  The lines
 at $45^{0}$ represent light rays. The speed of propagation of
 causation is limited by the speed of light. As indicated in the 
 figure, the region of causal contact between $t = 0$ and the
 time of recombination is smaller than the region of the last
 scattering surface (the intersection of our past light cone $l_p$
 with the surface at $t = t_{rec}$) over which the microwave background
 is observed to be isotropic. The maximal angular scale where isotropy
 can in principle be explained by causal physics is the angular scale
 of the first peak in the CMB angular power spectrum. 
 
 As Figure \ref{WMAPspect} shows, there is nontrivial structure in the CMB on angular
 scales which according to SBB cosmology could never
 have been in causal contact. Hence, SBB cannot explain the origin
 of structure on these large scales. This is the {\it formation of structure
 problem} from which the SBB scenario suffers.
 
 In addition, SBB cosmology cannot explain the degree of spatial flatness
 of the universe which is currently observed. According to SBB cosmology,
 a spatially flat universe is unstable to the development of curvature in the
 expanding phase. The near spatial flatness which is currently observed
 requires a tuning of the relative contribution of the spatial curvature to the total
 energy density which is of the order of $10^{-50}$ at energy scales corresponding
 to particle physics grand unification.
 
 Since the SBB scenario is well tested at late cosmological times, any
 solution of the above-mentioned problems will require new physics during
 the stages of the very early universe.
 
\subsection{Preview}

In the following Sections we will discuss different scenarios of very early
universe cosmology which can provide solutions to the horizon, flatness and
structure formation problems. We will compare the current paradigm,
the inflationary universe scenario, with a couple of alternatives. The
main message will be that there are a number of early universe scenarios,
and not just inflation, which are compatible with current observations.

We will then ask the question what kind of picture of the early universe
emerges if superstring theory is the correct theory which unifies all forces
of nature at high energies. Hints will be discussed which indicate that
superstring cosmology will be nonsingular and may not include a phase
of inflation.
  
\section{Theory of Cosmological Perturbations}
  
\subsection{Basics of Cosmological Perturbation Theory}

Observations of CMB anisotropies and large-scale structure concern
linear fluctuations about the cosmological background (\ref{bgcosmo}). 
Hence we must start with a brief summary of how these inhomogeneities
are described and how they evolve (see e.g. \cite{MFB, RHBpertrev}
for details). We here work in the context of Einstein gravity with
a matter source. For simplicity, 
matter is modelled in terms of a scalar field $\varphi$
with a non-trivial background dynamics $\varphi_0(t)$.
Since the universe is observed to converge to homogeneity
on large scales, the fluctuations can be described in linear theory.
This means that any fluctuating field can be expanded in
plane waves (Fourier modes), and each such mode evolves
independently. The Fourier modes can be labelled by the
comoving wavenumber $k$.

Linear fluctuations of geometry and matter can be
classified according to how they transform under spatial rotations.
Out of the ten degrees of freedom of the metric, four are scalars, four vectors,
and two tensors (the two polarization states of gravitational
waves). The physics is independent of which coordinates are used,
and hence there are four coordinate modes (gauge modes) which
can be factored out, leaving only two scalars and two vectors, plus
the gravitational waves. For simple matter models such as a 
scalar field with a homogeneous component which is evolving in
time (a {\it rolling} scalar field) the vector modes are not sourced at linear order in
perturbation theory, and hence we will not consider them. We
focus on the scalar modes which are sourced by matter. One of
the scalar modes vanishes for matter without anisotropic stress,
and the remaining mode is determined by the matter fluctuations.

We can choose coordinates in which the scalar metric fluctuations are
diagonal, and the metric is
\be \label{pertmetric} 
ds^2 = a^2 \left\{ \left(1 + 2 \Phi\right)
d\eta^2 - \left[\left(1 - 2\Phi\right) \delta_{ij} + h_{ij}\right] 
dx^i  dx^j \right\} , 
\ee 
where $\Phi$ is a function of space and time and represents
the scalar metric fluctuations. 
The matter field including linear fluctuations is
\be 
\varphi(\bm{x}, t) = \varphi_0(t) + \delta \varphi(\bm{x}, t) . 
\ee

The equations of motion for the fluctuations can be obtained by expanding 
the action of matter plus gravity to second order about the background. Since
the background satisfies the equations of motion, terms linear in
cosmological fluctuations cancel out in the action, leaving the
quadratic terms as the leading fluctuation terms. There is only
one canonical fluctuation variable $v({\bm{x}}, t)$ 
(variable in terms of which the action has a canonical kinetic term).
As shown in \cite{Sasaki, Mukh2}, this variable is 
\be 
v = a \left( \delta \varphi + \frac{z}{a} \Phi \right) , 
\ee 
where 
\be 
z \, = \, \frac{a \varphi_0^{\prime}}{{\cal H}} \, ,
\ee 
and obeys the equation of motion
\be  \label{vEoM}
 v_k^{\prime \prime} + \left( k^2 -
\frac{z^{\prime \prime}}{z} \right) v_k \, = \, 0 \, , 
\ee 
where a prime denotes the derivative with respect
to conformal time $\tau$. Here we see the crucial
role which the Hubble radius $H^{-1}$ plays. If
the equation of state of matter is constant in time,
then the factor $z^{\prime \prime} / z$ is of the order $H^2$.
Hence, on sub-Hubble scales ($k > H$) fluctuations
oscillate, while on super-Hubble scales ($k < H$) they
are {\it frozen in} and are {\it squeezed}, i.e. in an
expanding universe the amplitude of the dominant mode
of the above equation grows as $z \sim a$.

The curvature fluctuation $\zeta$ is in fact given by $v/z$.
What is measured is the power spectrum $P_{\zeta}(k)$ of $\zeta$
which gives the mean square fluctuation of $\zeta$ on a
length scale $k^{-1}$ and is given in terms of the Fourier
modes $\zeta(k)$ of $\zeta$ by
\be 
P_{\zeta}(k) \, = \, k^3 |\zeta(k)|^2 \, , 
\ee 
If this quantity is independent of $k$, one has a {\it scale-invariant}
spectrum. More generally, one has $P(k) \sim k^{n_s - 1}$ where $n_s$
is called the scalar spectral index. The observed spectrum is
nearly scale-invariant with a small red tilt \cite{Planck2}.

An initial vacuum spectrum at time $t_i$ is
\be
 \label{IC} v_k(t_\mathrm{i}) \, = \, \frac{1}{\sqrt{2 k}} \, .
 \ee 
This is a deep blue power spectrum, i.e. there is more
power on short wavelengths. If fluctuations originate as vacuum
perturbations (which they are assumed to in inflationary cosmology
and in a number of alternatives), then the squeezing of the
fluctuations on super-Hubble scales must have exactly
the right features to convert the initial vacuum spectrum to
a scale-invariant one. As will be discussed in the next
section, this happens both for inflationary cosmology and for
the matter bounce scenario. 

Note that the equation of motion for the Fourier mode of the
amplitude of gravitational waves is the same as (\ref{vEoM}) 
except that the function $z(\eta)$ is replaced by the scale factor
$a(\eta)$. If the equation of state of matter is constant, then
$z \sim a$ and hence the gravitational waves evolve as the
curvature fluctuations.
    
\subsection{Criteria for a Successful Early Universe Cosmology}
 
\begin{figure}
    \includegraphics[scale=0.65]{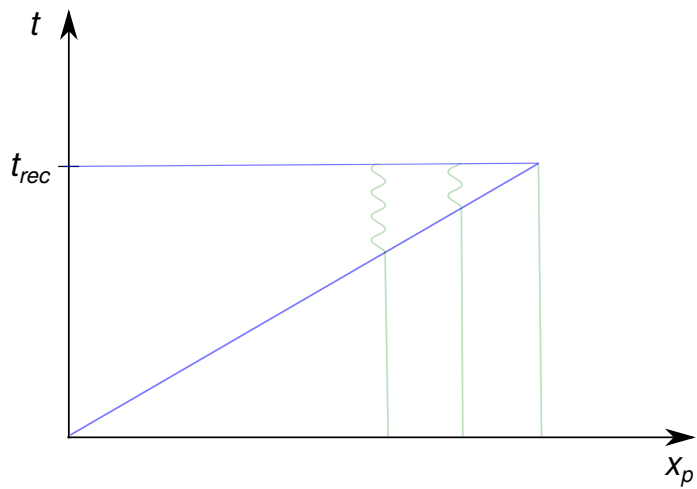}
    \caption{Origin of acoustic oscillations in the CMB angular power spectrum:
    assuming that fluctuations originate as standing waves on scales
    which are super-Hubble at early times, then different modes perform
    a different number of oscillations between when they enter the Hubble
    radius and start to oscillate and the time of recombination. Those which
    perform $n = 0, 2, 4, ...$ quarter oscillations are standing waves with
    maximal amplitude at $t_{rec}$ and yield peaks in the anisotropy spectrum,
    whereas those which perform $n = 1, 3, ..$ quarter oscillations only have
    velocity inhomogeneities at $t_{rec}$ and yield minima in the angular power
    spectrum (see \cite{SZ} from where the figure is adapted).}
    \label{SZfig}
  \end{figure}

The presence of acoustic oscillations in the angular power spectrum
of the CMB and in the power spectrum of matter density fluctuations
(the so-called {\it baryon oscillation peak}) were predicted \cite{SZ, PY}
a decade before the development of inflationary cosmology. In these
works it was realized that a roughly scale-invariant spectrum of
standing wave fluctuations on super-Hubble scales at cosmological
times before the time $t_{rec}$ of recombination would, as the scales
enter the Hubble radius and begin to oscillate as described in the previous
subsection, would be in good agreement with the data on the distribution
of galaxies which was available at the time when the papers were written,
and give rise to oscillations in the angular power spectrum of the
CMB and in the power spectrum of density fluctuations.  The dynamics
is illustrated in Figure \ref{SZfig}. At that time,
however, no models based on causal physics were known for how to
produce such primordial fluctuations. 
     
 Here we outline the criteria which a successful early universe scenario
 must satisfy.  Let us emphasize again the difference between the horizon and the
  Hubble radius. The horizon is the forward light cone of a point at
  the initial time and carries causality information. Beginning at the
  initial time, it is possible to have causal contact within the horizon.
  The Hubble radius $H(t)^{-1}$, on the other hand, is a local concept.
  It separates scales where fluctuations oscillate (sub-Hubble modes)
  from scales where the inhomogeneities are frozen in and get
  squeezed (super-Hubble modes).
  
  Since the Hubble radius at $t_{rec}$ is smaller than the radius of
  the part of the universe which we probe with the CMB, the condition
  for an early universe scenario to solve the horizon problem is
  that the horizon at $t_{rec}$ be much larger than the Hubble
  radius at that time (at least two orders of magnitude since the Hubble radius
  at $t_{rec}$ corresponds to an angular scale of about $1^{o}$ and we need
  to explain the absence of order one anisotropies on the angular scales
  of the full sky).
 Thus, the first requirement on a successful early universe scenario
 is that the horizon at late times be much larger than the Hubble radius.
 
 In order to be able to have a causal mechanism of structure formation,
 scales which are currently observed in cosmology must originate in
 the early universe with a wavelength smaller than the Hubble radius. This
 is the second requirement.
 
 Thirdly, the scales which we observe today must propagate for a long time
 at super-Hubble lengths. This is in order that the fluctuations are squeezed
 and enter the Hubble radius at late times as standing waves. This is
 required in order to provide an explanation for the observed acoustic
 oscillations in the angular power spectrum of the CMB.
 
 Finally, the generation mechanism which acts on sub-Hubble scales
 must yield a nearly scale-invariant spectrum of cosmological perturbations.
 
 The inflationary universe scenario provides a simple realization of these
 four criteria. However, there are alternative realizations as will be
 explained in the following section.

  \section{Paradigms of Early Universe Cosmology}
  
  \subsection{Cosmological Inflation}
  
  Cosmological inflation \cite{Guth, Brout, Sato, Fang} assumes that space underwent a
  period of almost exponential expansion during a finite time interval
  in the very early universe, starting at some time $t_i$ and ending
  at a time $t_R$ \footnote{The subscript ``R'' stands for {\it reheating}
  since at this time the conditions of a hot early universe must be
  created.}. The first model of inflation was based on a modified
  gravitational action \cite{Starob} where higher curvature terms dominate
  at early times and lead to accelerated expansion. Later models
  assumed that the accelerated expansion of space is realized in
  the context of Einstein gravity, but assuming the presence of a slowly
  rolling \cite{new1, new2, chaotic} scalar field whose energy-momentum tensor is
  dominated by the scalar field potential energy. After the end of
  inflation at time $t_R$, the evolution of the universe is like in Standard
  Big Bang cosmology.
  
\begin{figure}
    \includegraphics[scale=0.8]{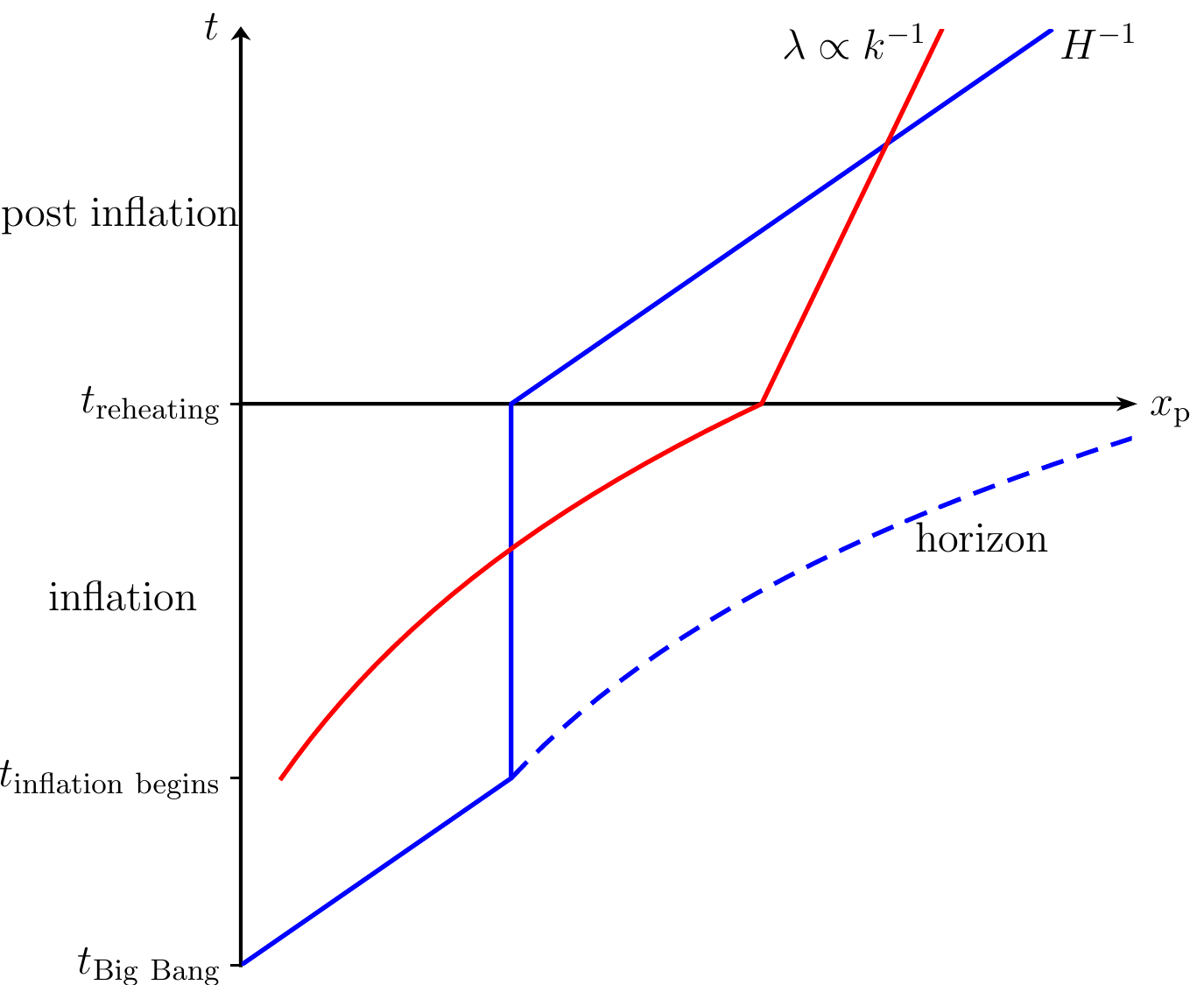}
    \caption
      {Space-time sketch of inflationary cosmology. The vertical axis is time, the horizontal
      axis is physical distance. The dashed curve represents the horizon, the solid blue
      line is the Hubble radius and the curve labelled by $k$ is the physical wavelength
      of the fluctuation on a particular (comovng) scale. Note that during the period inflation
      both the wavelength of a fluctuation and the horizon grow exponentially (for 
      exponential inflation) which is indicated by the curvature of the lines. After
      the end of inflation, the wavelength grows as $a(t) \sim t^{1/2}$ which is indicated
      by the near linearity of the curve.}
    \label{fig-infl}
  \end{figure}

  Figure \ref{fig-infl} represents a space-time sketch of inflationary cosmology. The
  vertical axis is time, the horizontal axis denotes physical spatial distance.
  The solid curve labelled $H^{-1}$ represents the Hubble radius, the
  dashed line is the horizon, and the curve labelled $k$ represents
  the physical wavelength of a fluctuation mode.
  
  During the phase of accelerated expansion for $[t_i < t < t_R]$, the horizon
  is growing very fast while the Hubble radius increases only slowly. For
  almost exponential expansion the horizon increases nearly exponentially
  while the Hubble radius is almost constant. Only a very short interval
  of accelerated expansion is needed to solve the horizon problem (the
  horizon needs to increase by a factor of greater than $100$ relative to
  the Hubble radius). 
  
  If the energy scale of inflation corresponds to the scale of particle physics
  Grand Unification (an energy scale of about $10^{16} {\rm GeV}$), then
  about 60 e-foldings of exponential expansion are required in order that
  scales which we observe now in cosmology had a wavelength smaller
  than the Hubble radius at the beginning of the inflationary period. In this
  case, the second of the general criteria for a successful early universe
  cosmology is satisfied. 
  
  As is evident from Figure \ref{fig-infl}, the wavelength of fluctuations is larger
  than the Hubble radius for a long time. Hence, the squeezing of the
  fluctuations which is required in order to obtain the acoustic oscillations
  in the CMB angular power spectrum is realized.
  
  Finally, as initially argued in \cite{Starob2} for gravitational waves and in
  \cite{ChibMukh} for the curvature perturbations, the power spectrum
  of fluctuations will be approximately invariant of scale. This is
  a consequence of the time-translation invariance of the phase of exponential
  expansion \cite{Press, Sato}. It is usually assumed that fluctuations originate as
  quantum vacuum perturbations at early times during the inflationary phase.
  This assumption can be justified since, in the absence of interactions with
  the agent driving the accelerated expansion, any initial fluctuations are
  redshifted away, leaving the matter in a vacuum state. If the agent driving
  inflation interacts with matter, it is possible to realize inflation while maintaining
  the dominance of thermal fluctuations. This is the {\it warm inflation}
  scenario \cite{Berera}.

The scale-invariance of the spectrum of curvature fluctuations can also
be seen using the general formalism described in the previous section.
The curvature fluctuation variable, starting out with a deep blue vacuum
spectrum on sub-Hubble scales, gets squeezed on super-Hubble scales.
Large wavelength fluctuations exit the Hubble radius earlier and are
thus squeezed more. For almost exponential inflation the squeezing is
just right to turn the initial vacuum spectrum into a scale-invariant one.
To see this, consider first the Hubble crossing condition
\be \label{crossing} 
a^{-1} (t_H t(k)) k \, = \, H \, . 
\ee 
Before $t_H(k)$, the amplitude of $v_k$ is constant, afterwards it
increases as $z$. Thus, the power spectrum of $\zeta$ at
some late time $t$ is 
\bea
P_{\zeta}(k, \eta) \, &=& \,  k^3 z^{-2}(\eta) \left(
\frac{z(\eta)}{z((\eta_H(k))} \right)^2 |v_k(t_i)|^2 \nonumber
\\
\, &\simeq& \, \frac{1}{2} \left(
\frac{a(t_H(k))}{z(t_H(k))} \right)^2 H^2
\eea
The $k$-dependence has cancelled out and we thus obtain a 
scale-invariant spectrum.

Gravitational waves evolve as curvature fluctuations, with the
function $z(\eta)$ being replaced by $a(\eta)$. Hence, inflationary
cosmology predicts a scale invariant spectrum of gravitational waves
whose amplitude is suppressed compared to that of curvature 
fluctuations by the slow-roll parameter $\epsilon$ which gives the
ratio between $z$ and $a$ during the inflationary phase.

  Of the early universe scenarios discussed here, inflationary cosmology is
  the only one which predicted (as opposed to post-dicted) various observational
  results, e.g. the spatial flatness of the universe, the almost scale-invariance
  of the spectrum of cosmological perturbations and the near Gaussianity
  of this spectrum. In fact, inflation predicted \cite{ChibMukh} a slight red
  tilt of the spectrum, a tilt which has now been established by the recent
  CMB observations (see e.g. \cite{Planck2}). Inflation predicts a nearly
  scale-invariant spectrum of gravitational waves \cite{Starob2} with a slight red-tilt. Note
  that, whereas the tilt of the spectrum of cosmological perturbations can
  be made blue by complicating the scalar field model of inflation, the
  spectrum of gravitational waves has a red tilt unless the matter which is
  responsible for inflation violates the energy condition $p + \rho \geq 0$,
  where $p$ and $\rho$ are pressure and energy density, respectively \footnote{At this point,
  we have not yet observed a stochastic background of gravitational waves, and 
  observing the small tilts predicted by various early universe models will be
  very challenging.}.
  
  A drawback of the inflationary scenario is the {\it trans-Planckian problem}
  for cosmological perturbations \cite{Jerome1, Jerome2} (see e.g. \cite{Jerome3}
  for a review with an extended list of references). If the period of inflation lasts
  only slightly longer than the minimal amount of time it needs to in order to
  solve the flatness and structure formation problems of SBB cosmology,
  then the wavelengths of the fluctuation modes which we observe today
  are smaller than the Planck length at the beginning of inflation. We do
  not understand the physics on these wavelengths, and hence it is unclear
  if the initial conditions for the fluctuations which are used are well justified.
  Some other problems of inflationary cosmology will be discussed in later
  sections.
  
  \subsection{Matter Bounce}
  
  Another paradigm to obtain successful structure formation is the
  {\it matter bounce} scenario \cite{Fabio, Wands}. It is assumed that
  there is new physics \footnote{The new physics could be string
  theory, loop quantum gravity or some effective field theory
  which corresponds to Einstein gravity coupled to some matter
  which violated the usual Hawking-Penrose energy conditions.
  For a review of bouncing cosmolgies see e.g. \cite{RHB-Peter}.
  We are emphasizing here ways to obtain a spectrum of
  cosmological fluctuations in agreement with observations, and
  the specific dynamics of the bounce phase has little to say about
  this issue.}
  which resolves the cosmological singularity.
  Time runs from $- \infty$ to $\infty$. The universe begins in a
  contracting phase which is the mirror inverse of the expanding
  Standard Big Bang cosmology evolution, which is to say that at
  very early times during contraction, a dark energy component may
  dominate, followed by a period of matter-dominated contraction,
  followed by a radiation-dominated phase, and then a nonsingular
  bounce. The bounce point (minimal value of the scale factor) is
  $t = 0$. 
  
   \begin{figure}
    \includegraphics[scale=0.8]{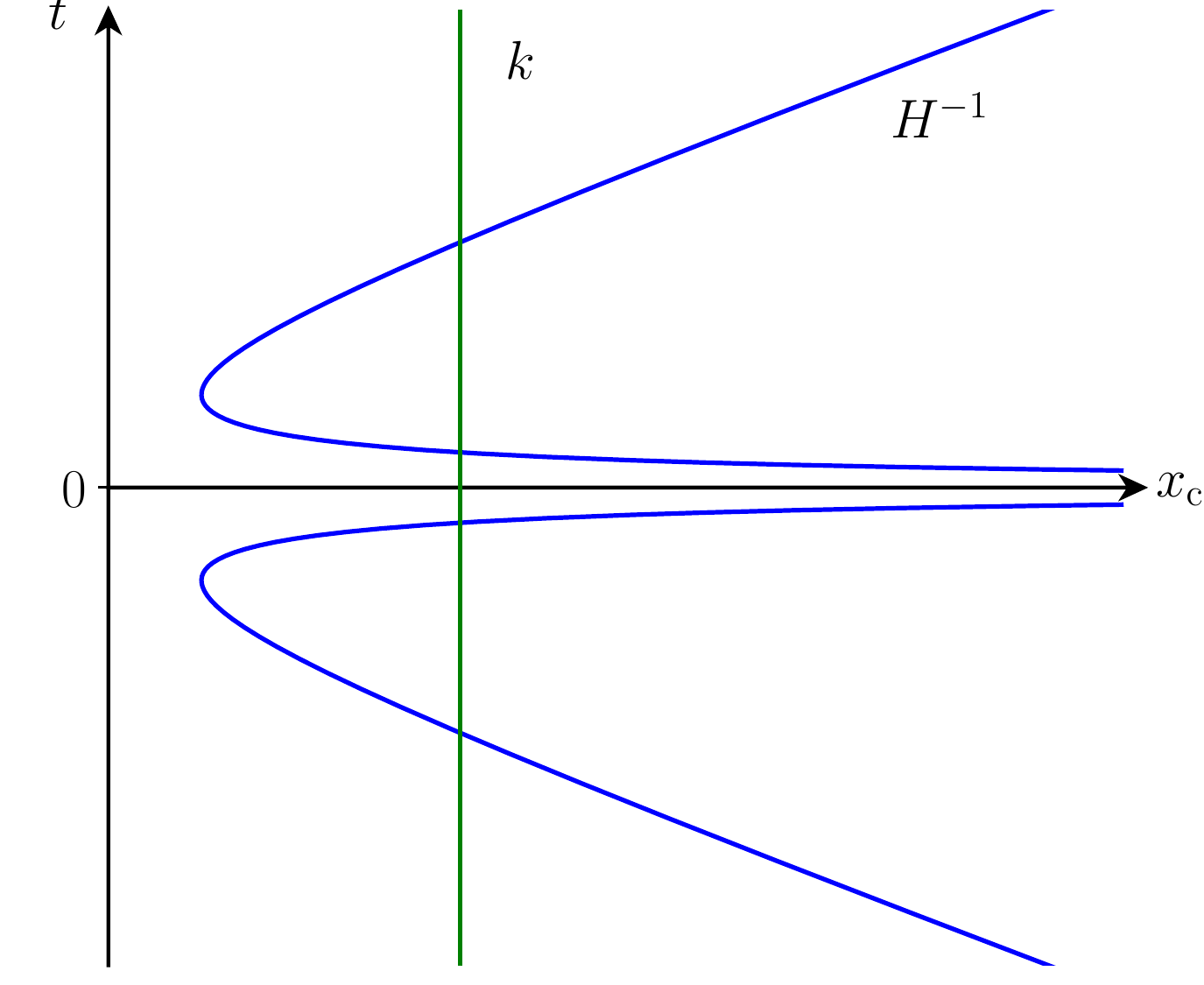}
    \caption{Space-time sketch of the matter bounce cosmology. The vertical
    axis is conformal time, the horizontal axis is comoving distance. The blue
    solid curve is the Hubble radius, the green vertical line represents the
    wavelength of a fluctuation mode. The bounce phase (for which new physics
    is required) lasts from the time when the Hubble radius takes on its minimal
    value in the contracting phase to the corresponding time during expansion. }
    \label{fig-bounce}
  \end{figure}
  
  Figure \ref{fig-bounce} presents a sketch of the resulting space-time diagram. The
  vertical axis is conformal time, the horizontal axis represents comoving spatial
  distance. The Hubble radius is symmetric about $t = 0$. 
  The perpendicular curve indicates the wavelength of a fluctuation.
  Since there is no origin of time, the horizon is infinite and there is
  no horizon problem. As obvious from the figure, all scales originate
  inside the Hubble radius, and hence a causal structure formation
  scenario is possible. In fact, for scales which exit the Hubble radius
  during the matter-dominated phase of contraction, an initial vacuum
  spectrum on sub-Hubble scales evolves \cite{Wands, Fabio} into
  a scale-invariant spectrum on super-Hubble scales. If we take
  into account the dark energy component in the contracting phase,
  a slight red tilt of the spectrum results, as in inflationary cosmology \cite{Ed}.
  
  The reason why a scale-invariant spectrum of curvature fluctuations
  is obtained from initial vacuum perturbations is that the squeezing
  function in (\ref{vEoM}) has the same dependence on $\eta$ as
  in the case of inflation. Hence, the conversion of a vacuum spectrum
  to a scale-invariant way proceeds as in the case of inflation.
  
  Note that a scale-invariant spectrum of gravitational waves is also
  generated in the matter bounce scenario. This demonstrates that
  the often stated claim that a detection of primordial gravitational
  waves on cosmological scales will confirm inflation is false (see
  \cite{gun} for a detailed discussion of this point). In fact, in the
  matter bounce scenario the analog of the inflationary slow-roll parameter
  $\epsilon$ is of order one, and hence there is no suppression of
  the amplitude of gravitational waves.
  
  The matter bounce scenario faces significant problems. For one,
  the contracting phase is unstable against anisotropies \cite{Peter}.
  In addition, there is no suppression of gravitational waves compared to
  cosmological perturbations, and hence the amplitude of gravitational
  waves is predicted to be in excess of the observational bounds.
 If the cosmological perturbations are boosted in the bounce phase,
 then non-Gaussianities are induced which are in excess of observational
 bounds \cite{Quintin}. Postulating a large graviton mass in the contracting
 phase can solve both of these problems \cite{Chunshan}.
   
  \subsection{Pre-Big Bang and Ekpyrotic Scenarios}
 
 The Pre-Big-Bang \cite{PBB} (PBB) and Ekpyrotic scenarios \cite{Ekp} are bouncing 
 cosmologies which avoids the anisotropy and overproduction of gravitational wave 
 problems of the matter bounce scenario.  In both of these cosmologies, the
 contracting phase is dominated by a form of matter whose energy density
 increases as fast (in the case of the PBB scenario) or faster (in the case of the
 Ekpyrotic scenario) than the contribution of anisotropies. In the case
 of the PBB scenario, it is the kinetic energy of the dilaton field (one of the
 massless degrees of freedom of string theory) which dominates in the
 contracting phase, in the case of the Ekpyrotic scenario it is the energy
 density of a new scalar field with negative exponential potential. Both of these
 scenarios were initially proposed based on ideas in superstring theory but they
 can also be viewed as effective field theories involving a new scalar field
 with some special features.
 
 The space-time sketches of the PBB and Ekpyrotic scenarios are similar to
 that of the matter bounce paradigm (see Figure \ref{fig-bounce}), except that the bounce is
 not symmetric. In particular, the horizon problem of Standard Big Bang
 cosmology is solved in the same way as in the matter bounce, and in
 the same way fluctuations on all scales observed today originate inside
 the Hubble radius at early times, thus allowing a causal structure formation
 scenario. However, unlike what happens in the matter bounce scenario,
 the growth of fluctuations on super-Hubble scales in the contracting phase
 is too weak to convert an initial vacuum spectrum into a scale-invariant
 one. The resulting spectrum of curvature fluctuations and gravitational
 waves is blue (see e.g. \cite{Fabio2}), thus not allowing initial vacuum
 perturbations to explain the observed structures in the universe, and
 predicting a negligible amplitude of gravitational waves on cosmological
 scales. As studied in \cite{PBBflucts} in the case of the PBB scenario,
 and in \cite{newEkp} in the case of the Ekpyrotic scenario, a scale-invariant
 spectrum of curvature fluctuations can be obtained by using primordial
 vacuum fluctuations in a second scalar field.
 
 \subsection{Emergent String Gas Cosmology}
 
 Another alternative to cosmological inflation as a theory for the origin
 of structure in the universe is the {\it emergent scenario} as realized
 in {\it String Gas Cosmology} (SGC) \cite{BV}. This scenario (see
 e.g. \cite{SGCrevs} for reviews) is based, as discussed in the following
 section, on the idea that there was a long quasi-static phase in
 the early universe dominated by a thermal gas of fundamental
 strings. This phase may be past-eternal, or it may be preceded
 by a previous phase of contraction. At the end of this phase there
 is a transition to the expanding radiation phase of SBB cosmology.
 
  \begin{figure}
    \includegraphics[scale=0.7]{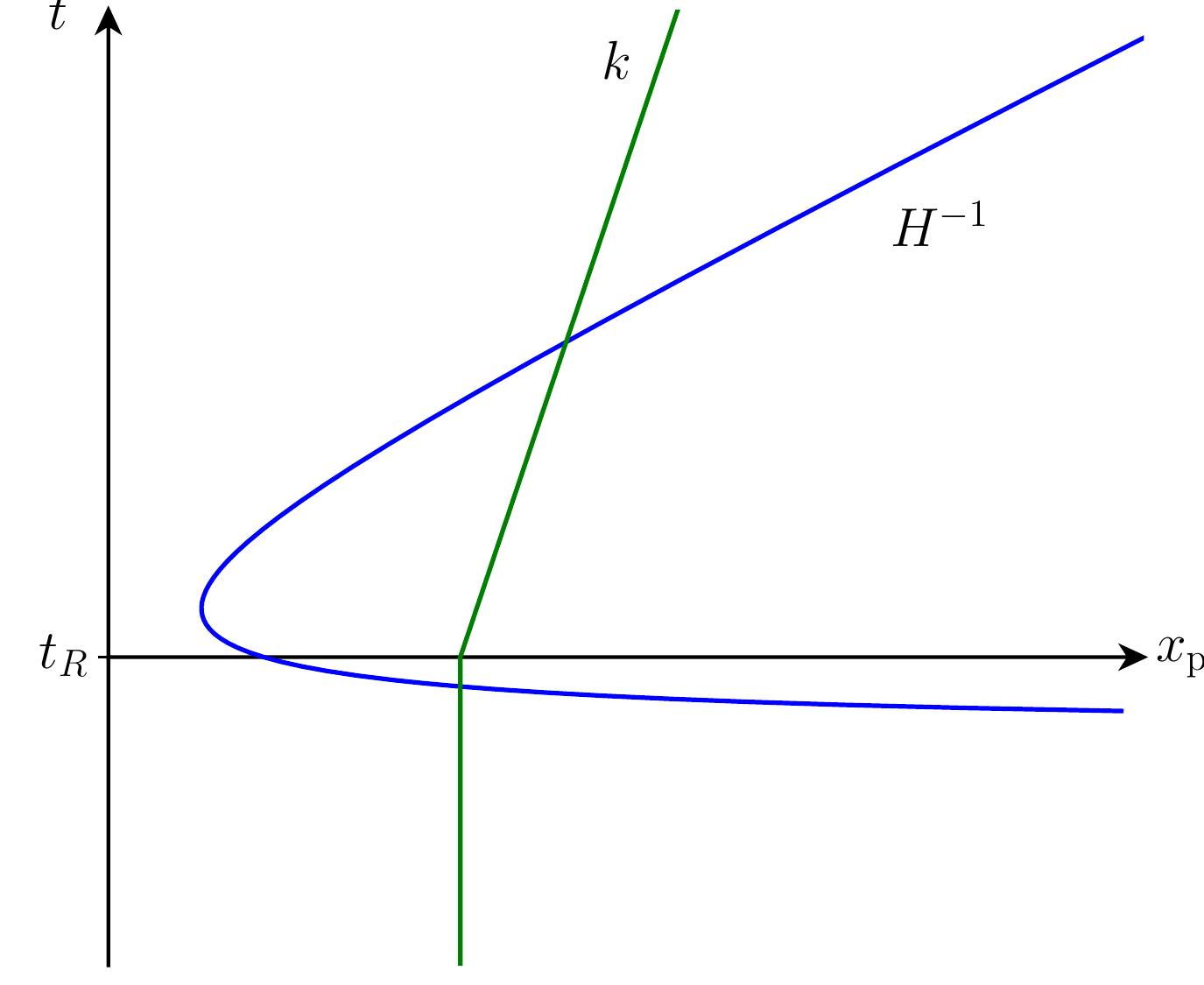}
    \caption{Space-time sketch of String Gas Cosmology. The vertical axis is
    time, the horizontal axis is physical distance. The time $t_R$ is the end
    of the Hagedorn phase and beginning of the radiation phase of SBB
    cosmology. The green curve labelled $k$ represents the wavelength
    of a fluctuation mode. The Hubble radius is labelled by $H^{-1}$.}
    \label{fig-sgc}
  \end{figure}
  
 Figure \ref{fig-sgc} is a space-time sketch of SGC. The vertical axis is
 time, with the time $t_R$ denoting the end of  the quasi-static
 phase, the horizontal axis is physical distance. The curve labelled
 $H^{-1}$ is the Hubble radius which is infinite in the Hagedorn phase, falls
 to a microscopic value at $t_R$ and then increases linearly in time
 as in SBB cosmology. Since time goes back to $- \infty$, there is
 no horizon problem, as in the bouncing models discussed in previous
 subsections. As is obvious from Figure \ref{fig-sgc}, all scales originate from
 inside the Hubble radius at early times, thus allowing for a causal
 structure formation scenario.
 
 As first realized in \cite{NBV, NBPV}, thermal fluctuations of a gas of closed
 strings on a compact space with the topology of a torus lead to a
 scale-invariant spectrum of curvature fluctuations and gravitational
 waves. The tilt of the spectrum of curvature fluctuations is predicted
 to be red (as in the case of inflationary cosmology), but that of the
 gravitational waves is slightly blue, in contrast to what is obtained
 in inflation. The amplitude of gravitational waves is suppressed
 compared to that of curvature fluctuations by the equation of state
 parameter $w = p/\rho$ \cite{NBPV}.
  
The emergent string gas cosmology model is an example where
the fluctuations are of thermal origin, not of quantum origin as
they are in most inflationary models (warm inflation being the
exception) and the Ekpyrotic scenario.
  
\subsection{Comparisons}

 The main messages from this section is that there are a number of
 different early universe scenarios which can solve the horizon problem
 of SBB cosmology and which can generate a spectrum of curvature
 fluctuations consistent with observations. Inflationary cosmology is
 not the only game in town. One must, however, admit that
 inflation is the first scenario proposed and the only one
 which can claim to have made successful {\it predictions} (as opposed
 to post-dictions). Inflation is also the only scenario which is at the
 present time self-consistent from the point of view of low energy
 effective field theory (Einstein gravity coupled to low energy quantum
 field matter). All the other scenarios mentioned above need new
 physics to obtain essential aspected of the dynamics of the cosmological
 background, be it a cosmological bounce or a quasi-static phase.
 Another nice feature which inflation has is that - at least in the
 case of large-field inflation - the slow roll trajectory in field space
 which leads to inflation is a local attractor in initial condition space
 \cite{Kung, RHBinflRev}. This feature is shared by the Ekpyrotic
 scenario, but not some of the other ones.
 
 On the other hand, in all of the alternatives to inflation mentioned
 above, the physical wavelength of the fluctuations which are currently
 observed are much larger than the Planck length (as long as the
 maximal temperature is smaller than the Planck scale) at all times.
 Hence, the trans-Planckian problem which the theory of
 cosmological fluctuations in inflationary cosmology suffers from is
 not present.
 
 Unless model parameters are finely tuned, the energy
 scale at which inflation takes place is close to the particle physics
 scale of {\it Grand Unification}. This is close to the Planck scale
 and even closer to the preferred string scale \cite{GSW}. Hence,
 the extrapolation of low energy physics to the scale of inflation and
(to the bounce scale in bouncing cosmologies) needs to be justified.
In spite of a large body of work, there have so far not been any
convincing realizations of inflation in the context of superstring theory,
and there are indications that it might not be possible to embed simple
inflation models in string theory \cite{Vafa}.
Hence, there are good reasons to look beyond standard inflationary
cosmology.

Note that the structure formation scenarios described above are
not the only ones. The main goal was to present a few very different
scenarios and to show how the general criteria for a successful
early universe model can be realized.
          
\section{Hints from Superstring Theory}
  
\subsection{Challenges}

The goal of superstring theory is to provide a quantum theory which unifies all four
forces of nature, including gravity (see e.g. \cite{GSW, Pol1, Pol2} for
textbook treatments of string theory). If nature is indeed described by superstring
theory, then string theory should play a crucial role at the high energy densities which
were present in the very early universe. Whichever of the scenarios for structure
formation described in the previous section is in fact realized in nature should then
be determined by string theory. 

It has been shown to be very challenging to obtain an inflationary phase
from string theory (see e.g. \cite{Liam} for a detailed discussion). The problem
is that in order to obtain a period of slow-roll inflation from simple scalar field
potentials, field values in excess of the Planck mass $m_{pl}$ are required
(see e.g. the review in \cite{RHBinflRev}). However, for such large field values
string effects on the shape of the potential need to be considered, and tend
to destroy the required flatness of the potential unless special symmetries
of the field theory (e.g. shift symmetry \cite{Yama}) are considered. But even
in this case string theory arguments such as the {\it Weak Gravity Conjecture}
\cite{WGC} tend to invalidate the effective field theory constructions.

\subsection{String Thermodynamics}

\begin{figure}
    \includegraphics[scale=0.55]{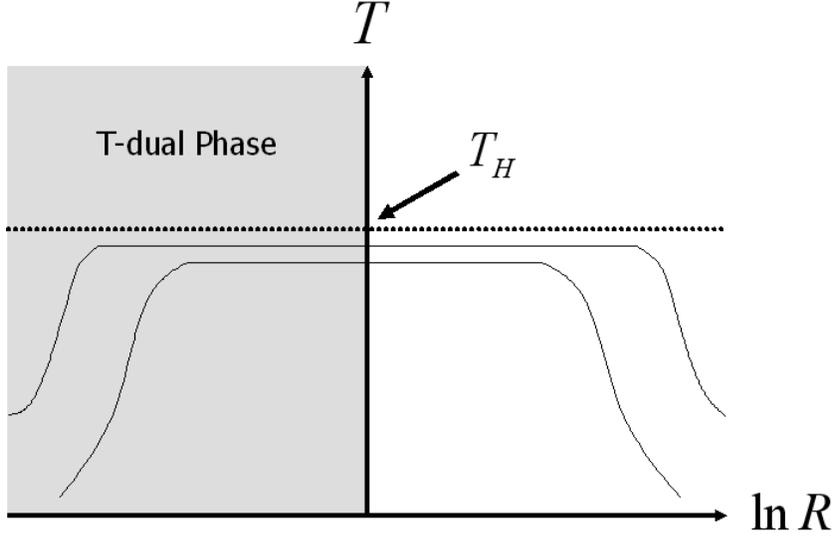}
    \caption{Temperature (vertical axis) as a function of the radius (horizontal axis) of a box
    of strings in thermal equilibrium.}
    \label{cantor}
  \end{figure}

There are indications that the description of the very early universe in string theory
will look very different from what can be obtained by models based on point particle
theories (which includes all existing ``string-derived'' low energy effective field
theories). The first evidence for this comes from string thermodynamics. It has
been known \cite{Hagedorn} from the earliest days of string theory that there is
a maximal temperature $T_H$ of a gas of closed string in thermal equilibrium, the so-called
{\it Hagedorn temperature}. The value of this limiting temperature is given by the
string scale. Assuming that space is a torus of radius $R$, then the evolution of the
temperature $T$ as a function of $R$ is sketched in Figure \ref{cantor}. The length of
the plateau region of $T(R)$ depends on the total entropy of the system: the larger
the entropy, the larger the plateau region.

The origin of the $T(R)$ curve of Figure \ref{cantor} is easy to understand. On a compact
space, there are three types of states of strings: momentum modes which correspond
to the centre of mass motion of the string, winding modes (which count the number
of times the string wraps the torus), and the string oscillatory modes (whose energies
are independent of $R$). The number of oscillatory states increases exponentially
with the energy of the state. It is this fact which leads to the existence of the limiting
temperature. As the energy density in a system is increased, then the energy
will go into the excitation of new oscillatory states as opposed to the increase in
the temperature.

The energies of the momentum and winding modes
are quantized in units of $1/R$ and $R$, respectively. For momentum modes we have
\be
E_n \, = \, n \frac{1}{R} \, ,
\ee
and for winding modes 
\be
E_m \, = \, m R \,
\ee
(in string units), where $n$ and $m$ are integers. In fact, there is one momentum
and one winding quantum number for each compact spatial dimension.

The above is the mass spectrum of free string states. It reflects an important
symmetry of string theory, the T-duality symmetry which implies that under the
transformation
\be
R \, \rightarrow \, \frac{1}{R}
\ee
(in string units) the spectrum of states is unchanged. This symmetry corresponds to
an interchange between momentum and winding quantum numbers. It is a symmetry
of string interactions, and it is assumed to be a symmetry of non-perturbative string
theory, giving rise to the existence of D-branes \cite{Pol2} (see also \cite{Timon}).
This symmetry allows us to understand the symmetry of the $T(R)$ curve of
Figure \ref{cantor} about ${\rm ln} R = 0$: for large values of $R$ the energy wants to
be in the modes which are light for large values of $R$, namely the momentum
modes, while for $R \ll 1$ in string units the energy will drift into the winding modes
which are the light one in this range of $R$ values.

Figure \ref{cantor} immediately \cite{BV} leads one to expect that in the context of
string theory the cosmological singularity can be avoided
\footnote{The logic here is that if the temperature remains finite, the curvature
should not be able to blow up.}, while in the context
of Einstein gravity coupled to particle matter a temperature singularity as
$R \rightarrow 0$ is unavoidable.

\subsection{T-Duality Symmetry and Doubled Space}

What is missing from the previous discussion is the dynamics of the background
space-time. This can obviously not be described by the Einstein-Hilbert action since
the this action is incompatible with the T-duality symmetry of string theory.
Dilaton gravity as described in \cite{PBB} is a better starting point. In this case
the T-duality symmetry of string theory is reflected in the so-called {\it scale factor duality}
which involves a transformation of both the metric and the dilaton field.

However, there may be a modified background description which is more
useful for superstring cosmology. The starting point of this description is the 
following: In quantum mechanics the position eigenstates $|x \rangle$ are dual to momentum eigenstates 
$|p \rangle$. In a toroidal background, the momenta are discrete, labelled by integers $n$, 
and hence
\be
|x \rangle \, = \, \sum_n e^{i n x} |n \rangle \, .
\ee
where $|n \rangle$ are the momentum eigenstates with momentum quantum numbers $n$.
As discussed above, in string theory on a torus the
windings are T-dual to momenta, and it is possible \cite{BV}
to define a T-dual position operator
\be
|{\tilde x} \rangle \, = \, \sum_m e^{i m {\tilde x}} |m \rangle \, ,
\ee
where $|m>$ are the eigenstates of winding, labelled by an integer $m$.

String states with both momenta and windings can be viewed as point
particles propagating on a {\it doubled space} which is spanned by both the
position and the dual position eigenstates. The coordinates of this space
are $X^{i}$
\be
 X^{i} \, = (x^{i}, {\tilde x}^{i}) \, .
 \ee
 If the underlying target space of the strings has $d$ spatial dimensions,
 the number of spatial dimensions of the doubled space is $2d$. This is
 the same space which is used in the {\it Double Field Theory} (see \cite{DFT}
 and \cite{DFTrev} for a review) approximation to string theory.
 
\begin{figure}
    \includegraphics[scale=0.8]{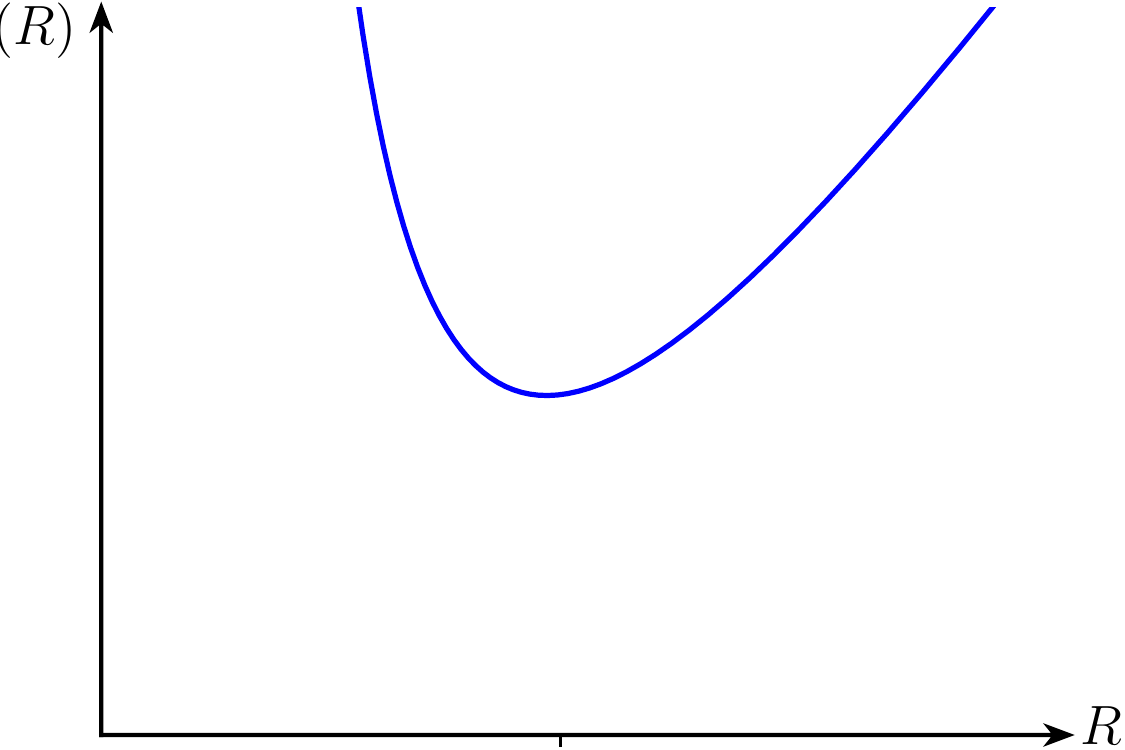}
    \caption{Physical length (vertical axis) as a function of the coordinate length (horizontal
    axis).}
    \label{phys_length}
  \end{figure}

 Let us consider a torus with radius $R$. If $R$ is large in string units,
 then the light states are the momentum modes and a physical apparatus
 to measure length will be built from momentum modes. However, if $R$
 is small compared to the string length, then it is the winding modes which
 are light and hence a physical apparatus will be constructed from the
 winding modes. Hence, an apparatus measuring the physical
 length $l_p(R)$ will be measuring (see Figure \ref{phys_length})
 \bea
 l_p(R) \, &=& \, R \,\,\,\,\, R \gg 1 \\
 l_p(R) \, &=& \, \frac{1}{R} \,\,\,\,\, R \ll 1 \, , \nonumber
 \eea
 when $R$ is expressed in units of the string length \cite{BV}.
 
 The above yields a new interpretation of a dynamical evolution
 $R(t)$. The physical interpretation of $R$ decreasing towards $R = 0$
 is that the dual radius (which is also the physical radius) is
 increasing to $R = \infty$ in the dual directions. In double space
 we can express this dynamics by taking a cosmological metric
 in double space which is
 \be
 ds^2 \, = \, dt^2 - a(t)^2 dx^2 - a(t)^{-2} d{\tilde x}^2 \, .
 \ee
 As mentioned above, physical measuring sticks measure length in terms
 of the coordinates related to the light string modes. Hence, for $R > 1$
 (in string units), length will be measured in terms of $x$, while for
 $R < 1$ it is measured in terms of ${\tilde x}$. Thus, a physical
 device will see space as contracting as $R$ decreases towards
 $R = 1$, but for $R < 1$ it will be seen as increasing. Thus, a
 physical observer will see no singularity.
 This argument is elaborated on in a recent paper \cite{Guilherme}.

 \section{Discussion}
  
In spite of the fact that the inflationary scenario has been a model of early
universe cosmology with many successes, we may have to look beyond
the standard inflationary scenario in order to understand the complete
evolution of the early universe. All inflationary practitioners would admit
that standard inflation cannot explain the evolution all the way back
to the Big Bang as a consequence of the cosmological singularities which
it does not avoid \cite{Borde}. Although it is possible that the correct
picture of the very early universe will involve a new phase followed
by a period of inflation like we understand inflation today, this needs
not be the case: there are a number of alternative early universe
scenarios which lead to cosmological structure formation consistent
with current observations.

Of all of the structure formation paradigms, inflation may be the
only one which is self-consistent at the level of effective point
particle field theories. However, if nature is described by string
theory, then it may be difficult to embed standard inflation into
the model, and an alternative such as string gas cosmology may
emerge in a more natural way, as already argued in \cite{BV}.
 
 \section*{Acknowledgement}
\noindent

The final section of this article is based on collaborative research with
Renato Costa, Guilherme Franzmann and Amanda Weltman, research
which is supported by the IRC- South Africa - Canada Research Chairs
Mobility Initiative Grant No. 109684. I would like to express my thanks to all of my
collaborators on the research reported on here. In particular, I
thank Elisa Ferreira for producing many of the figures, and Guilherme Franzmann
for comments on the draft. I also thank the 
Canadian NSERC and the Canada Research Chair program for partial
financial support.



\end{document}